\parindent=0cm

\font\gross=cmr10 scaled \magstep1

\font\mengen=bbm10

\font\csczwoelf=cmcsc10 scaled \magstep1

\font\ninebf=cmb7

\def\datum{\line{\hfil Berlin, den \the\day.\the\month.\the\year}}
\def\date{\line{\hfil Berlin,  \the\month/\the\day/\the\year}}

\def\doppel{\magnification=\magstep1}

\def\titel#1#2{{\removelastskip\bigskip\goodbreak\noindent\mark{#1}\gross
#1\medskip \nobreak\noindent #2\unskip}}
\def\subtitel#1#2{{\removelastskip\bigskip\goodbreak\noindent\bf
#1\medskip \nobreak\noindent #2\unskip}}
\edef\ignore#1{}

\def\lrkopfzeile{\nopagenumbers\headline={\ifodd\pageno\ungeraderkopf\else
\geraderkopf\fi}
\voffset=2\baselineskip}
\def\geraderkopf{\rm\folio\ \dotfill\ \it\firstmark}
\def\ungeraderkopf{\it\firstmark\ \dotfill\ \rm\folio}

\def\Tr{\hbox{Tr}}
\def\tr{\hbox{tr}}
\def\part#1#2{{\partial #1\over\partial #2}}
\def\br#1#2{{#1\over #2}}
\def\frac#1#2{{#1\over #2}}

\def\sla#1{\hbox to 0cm{/\hss}#1}

\def\MZ{\hbox{\mengen Z}}
\def\MR{\hbox{\mengen R}}
\def\MC{\hbox{\mengen C}}

\def\cS{{\cal S}}


\def\lrvec#1{\vbox{\ialign{##\crcr$\leftrightarrow$\crcr\noalign{\kern-1pt\nointerlineskip}$\hfil\displaystyle{#1}\hfil$\crcr}}}
\def\pfeilrmittext#1{\setbox3=\hbox{#1\kern
0.5em}$\displaystyle\mathrel{\mathop {\hbox to
\wd3{\rightarrowfill}}^{\box3}}$}  
\def\pfeillmittext#1{\setbox3=\hbox{\kern
0.5em #1}$\displaystyle\mathrel{\mathop {\hbox to
\wd3{\leftarrowfill}}^{\box3}}$}  
\def\pfeilrmittextou#1#2{\setbox3=\hbox{#1\kern
0.5em}$\displaystyle\mathrel{\mathop {\hbox to
\wd3{\rightarrowfill}}^{\box3}_{\hbox{#2\kern 0.5em}}}$}  
\def\pfeillmittextou#1#2{\setbox3=\hbox{\kern
0.5em #1}$\displaystyle\mathrel{\mathop {\hbox to
\wd3{\leftarrowfill}}^{\box3}_{\hbox{\kern 0.5em #2}}}$}  
%

%

\def\putlogo{\vbox to 0cm{\hbox to
0cm{\noindent\line{\hss\epsfbox{AEIlogo.eps}}\hss}\vss}}
\def\Tr{\mathop{\hbox{Tr}}}

\newcount\figcount \figcount=0
\def\fig#1#2{\global\advance\figcount by1\midinsert\vskip #2
\centerline{Fig.\ \the\figcount : #1}\endinsert}
\def\figstuff#1#2{\global\advance\figcount by1\midinsert #2
\centerline{Fig.\ \the\figcount : #1}\endinsert}
\def\nextfig{Fig.~{\advance\figcount by 1\relax\the\figcount}}
\def\psfig#1#2{\global\advance\figcount
by 1 \midinsert\vbox{\centerline{\epsfbox{#2}}
\centerline{Fig.\ \the\figcount : #1}}\endinsert}
\def\rpsfig#1#2#3{\global\advance\figcount
by 1 \xdef\#3{\the\figcount}\midinsert\vbox{\centerline{\epsfbox{#2}}
\centerline{Fig.\ \the\figcount : #1}}\endinsert} 

\def\ez#1e#2{#1\cdot 10^{#2}}
\def\crossout#1{\hbox to 0cm{\raise 1ex \hbox{$\underline{\hbox{\phantom{#1}}}$}\hss}#1}
\def\foryoureyesonly#1{}
%
\def\paper#1#2#3#4#5#6#7#8#9{\item{\bf [#1]}#2: ``#3'', #4 {\bf
#5} (#6) p.\ #7 #8 \foryoureyesonly{#9}\par\goodbreak}
%
%
\def\buch#1#2#3#4#5#6#7#8{\item{\bf [#1]}#2: ``#3'' (#4) #5 #6 #7 \foryoureyesonly{#8}\par}
%
%
\def\eprint#1#2#3#4#5#6{\item{\bf [#1]}#2: ``#3'', {{\tt #4}} #5
\foryoureyesonly{#6}\par}
%
%

%
%
\newcount\chapterno
\newcount\subchapterno
\newcount\glno
\edef\rememberref#1{\relax}  
\def\rememberchapter#1{\relax}  
\def\remembersubchapter#1{\relax}  
\def\neueseitevorchapter{\vfill\supereject\ifodd\pageno\relax\else \noindent$
$\vfill\eject\fi} 
\def\kapitelnummer{\the\chapterno}
\def\chapter#1{\neueseitevorchapter\global\advance\chapterno by
  1\expandafter\rememberchapter{\the\chapterno . #1}\glno =0
  \subchapterno =0\titel{\the\chapterno . #1}} 
\def\subchapter#1{\global\advance\subchapterno by
    1\remembersubchapter{\the\chapterno .\the\subchapterno
    . #1}\subtitel{\the\chapterno .\the\subchapterno . #1}}
\def\gln#1{\global\advance\glno by 1\xdef#1{(\the\chapterno
.\the\glno)}\eqno {(\the\chapterno .\the\glno)}}
\def\egln#1{\global\advance\glno by 1\xdef#1{(\the\chapterno
.\the\glno)}& {(\the\chapterno .\the\glno)}}

\def\openbib{\def\aux{1}\openout\aux=\jobname.aux \immediate\write16{Putting
    references in \jobname.aux}
    \def\rememberref##1{\write\aux{cite(##1) on page \folio}}
    \def\rememberchapter##1{{\edef\x{\noexpand\write\aux{chapter[##1] on page
    \noexpand\folio}}\x}}
    \def\remembersubchapter##1{\write\aux{subchapter[##1] on page \folio}} }
\def\closebib{\closeout\aux \def\rememberref##1{\relax}}
\def\cite#1{\raise 0.5ex\hbox{\ninebf[#1]}\rememberref{#1}}

\newif\ifpdf
\ifx\pdfoutput\undefined\pdffalse\else\pdfoutput=1\pdftrue\fi

{\catcode`\%=11\catcode`\!=14
\gdef\bluebg{\pdfliteral{
}
\def\aeiseminar{\magnification = 2500\raggedright
        \ifpdf\pdfpagewidth=35true cm\hsize=
        \pdfpagewidth  
        \bluebg
        \BrickRed\else\textBrickRed\fi}
\def\color{\ifpdf\input pdfcolor\else\input colordvi\fi}



\input epsf
\doppel
\openbib
\def\neueseitevorchapter{\par}
\parskip=1ex plus 1ex minus 0.5ex
\def\O#1{{\cal O}(#1)}
{\nopagenumbers
\line{\hss HU-EP-01/49}
\line{\hss hep-th/0111077}
\vskip 0.3\vsize plus 0.15 fil minus 3cm
\centerline{\csczwoelf A Remark On Field Theories On The Non-Commutative
Torus} 
\vskip 4cm plus 0.4fil
\centerline{Robert C. Helling\footnote{$^1$}{{\tt helling@AtDotDe.de}}}
\centerline{Institut f\"ur Physik}
\centerline{Humboldt-Universit\"at zu Berlin}
\centerline{Invalidenstra\ss e 110}
\centerline{D-10115 Berlin}
\centerline{Germany}
\vskip 2cm plus 0.4fil
{\bf Abstract:}

We investigate field theories on the non-commutative torus upon varying
$\theta$, the parameter of non-commutativity. We argue that one should think
of Morita equivalence as a symmetry of algebras describing the same space
rather than of theories living on different spaces (as is 
T-duality). Then we give arguments why physical observables depend on $\theta$
non-continuously.
\vskip 3cm plus 0.1fil
\eject}
\chapter{Introduction}
Non-commutative geometry, a generalization of ordinary geometry to spaces
with coordinate functions that generate non-commuting algebras, was pioneered
by Connes and has been studied for a long time, see\cite{C}, for a review also
\cite{L}. There is the general belief that a theory of quantum gravity
has to incorporate this generalization since Heisenberg's uncertainty relation
together with the Schwarzschild relation between mass and radius of a black
hole hint to a space-space uncertainty relation in a theory that incorporates
both, quantum physics and general relativity.

More recently, it was realized by Douglas and coworkers\cite{CDS}\cite{DH}\ 
that string theory, the leading candidate for a theory of quantum gravity,
indeed realizes non-commutative structures of the form introduced earlier in a
pure mathematical context. Since that time and especially after the work of
Seiberg and Witten\cite{SW}\ on non-commutative geometry in string theory, a
tremendous number of papers have been published that deal with that
subject. For a review consult\cite{DN}.

In string theory, non-commutativity enters thru the vev of the Neveu-Schwarz
two-form $B_{\mu\nu}$ that is the anti-symmetric cousin of the metric. In the
simplest case in which the $B$ field is just constant, the pointwise product
of functions is deformed to the $*$-product
$$(f*g)(x) = e^{\br i2 \theta^{\mu\nu}{\partial\over\partial
y^\mu}{\partial\over\partial z^\nu}}f(y)g(z)\bigg|_{x=y=z}$$
where $\theta$ is some function of $B$.

The nature of this background field entering as a deformation suggests that
also physical observables are just continously deformed from their values in
the commutative theory. Here however we will meet a surprise: In the
classical case a theorem by Gelfand and Naimark states that there is a one to
one correspondence between Hausdorff spaces and commutative C*-algebras via the
spectrum, that is the algebra of functions on that space. This is no longer
true in the non-commutative case: There, the mapping is one to many: There are
several different (more precisely: unitary inequivalent) algebras that
correspond to the same space. Thus, rather than considering single algebras as
representatives of spaces one should group them into equivalence classes. This
is the origin of Morita equivalence.

Our main example in this note will be the non-commutative two torus. Here, the
*-product is given in terms of a single parameter $\theta=\theta^{12}$. It is
well known that the Morita equivalence classes are orbits of the $SL(2,\MZ)$
action
$$\theta\mapsto{a\theta+b\over c\theta+d},\qquad a,b,c,d\in\MZ,\qquad
ad-bc=1,$$ 
that is for rational $\theta=p/q$:
$$\pmatrix{p\cr q\cr}\mapsto \pmatrix{a&b\cr c&d\cr}\pmatrix{p\cr q\cr}.$$
Thus in arbitrary small neighborhoods of two values of $\theta$ there are
always values that are Morita equivalent and that therefore describe the same
space. 

Therefore it seems highly unlikely that properties of the non-commutative
space and thus physics of field theories living on those spaces varies
continously as one varies $\theta$. This dependence of physics on $\theta$ is
the subject of this note. It has been studied from other perspectives in
\cite{GT}\ and \cite{AGB}.

The structure of this note is as follows: In the following section we discuss
Morita equivalence more extensively. In section three we show how it acts on
field theories on non-commutative spaces. The next section deals with
solutions of a toy ``equation of motion'' for different values of $\theta$. We
show that observables indeed vary discontinuously. The final section sets this
in the context of string theory and discusses the relation to the large volume
limit of the non-commutative plane.

{Acknowledgments: We have benefitted a lot from discussions with Luis
Alvarez-Gaume, Dieter L\"ust, Karl-Henning Rehren, Andreas Recknagel, Zachary
Guralnik, and Ralph Blumenhagen. This work is supported by the Deutsche
Forschungsgemeinschaft.}

\chapter{Morita equivalence of algebras describing a space}
In many situations it is fruitful in order to study an object $X$ to adopt a
dual perspective, that is to consider the algebra of morphism from $X$ to the
complex numbers. This, for example is subject of non-commutative geometry
(where $X$ is a topological Hausdorff space, perhaps endowed with additional
structure like differentiable structure, metric etc.), of quantum groups (where
$X$ is a Lie group) or quantum information theory (where $X$ is the classical
state space of a physical system operating on information).

In this note, we will be concerned with non-commutative geometry which in its
simplest form is about the Gelfand-Naimark isomorphism of the
categories of topological Hausdorff spaces and abelian C* algebras via the map
$$X \in \{\hbox{Hausdorf spaces}\} \mapsto {\cal A}_X :=\{f\colon X\to\MC | f
\hbox{ continuous, vanishing at $\infty$}\}$$

An important aspect of this correspondence is how to go in the opposite
direction, that is given an abelian C* algebra $\cal A$, construct the
topological space such that its algebra of continuous functions is given by
$\cal A$. The first step in this direction is to recover what is the analog
of a point  $x\in X$ in the algebraic setting. It is well known (for a review,
see for example \cite{L}) one can associate two structures to $x$: The
first is an irreducible representation of $\cal A$, as
$$\pi_x\colon{\cal A}\to\MC,\qquad \pi_x(f) := f(x)$$
is one and all of them are of this form. Furthermore, associated to $x$ there
is a maximal ideal in $\cal A$, that is the vanishing ideal of $\pi_x$, namely
$$\{f\in{\cal A}|f(x)=0\}. $$ 
Again, there is a one to one to one correspondence between maximal ideals,
points and irreducible representations, at least as long as $\cal A$ is
abelian.

If one now leaves the commutative realm and generalizes the above notions to
non-commutative C* algebras in general, these notions do not coincide
anymore\cite{L}\ and it is ambiguous what one should take as the
non-commutative extension of the inverse Gelfand-Naimark map, that is what
generalizes to the points of a non-commutative space that constitute the
space. 

Nevertheless, it is clear the structure of the space is --- as in the
commutative case --- encoded in the representation theory of the algebra. But
quite different from the commutative case where the representation theory is in
one to one correspondence with topological Hausdorff spaces, there are
inequivalent C* algebras with identical representation theory. 

Thus, one should not take one algebra to be a description of a non-commutative
space but rather an equivalence class of algebras. The appropriate equivalence
is given by Morita equivalence. Two (unitary) inequivalent algebras that are
nevertheless Morita equivalent should be thought of as two different
coordinatizations of the same non-commutative space.

To precisely define Morita equivalence one needs the  algebra $\cal K$ of
compact operators on a Hilbert space, that is the norm closure of finite rank
operators (in more physics parlance: All operators that can be obtained as
limits of series of matrices with only a finite number of non-zero
entries). Then, two algebras $\cal A$ and $\cal B$ are said to be Morita
equivalent if ${\cal A} \otimes {\cal K}$ is equivalent to ${\cal B} \otimes
{\cal K}$ in the ordinary (that is unitary) sense. (For an introduction see
\cite{L}\ or \cite{S}). Note that, so far, we have been concerned
only with the geometry of the space and have not mentioned field theories
living on the space.

As an example, note that $\cal A$ is Morita equivalent to the algebra
$M(n\times n,{\cal A}) = {\cal A} \otimes M(n\times n,\MC)$ of $n \times n$
matrices with entries in $\cal A$.

The fact that Morita equivalent algebras also have the identical K-theory
(which encodes the topology and possible stable branes in string theory) also
supports the view that all algebras in  an Morita equivalence class should be
thought of as the same non-commutative space.

The next step is to consider field theories defined on these non-commutative
spaces. If two algebras really describe the same space they should also
support the same field theories. This is really the main motivation for this
letter: One should think of Morita equivalence not as a property of field
theories living on a non-commutative space but rather as a property of the
algebras describing the space itself.

An alternative characterization of Morita equivalence is via the notion of
Hilbert bi-modules. Take again two algebras $\cal A$ and $\cal B$, then a
Hilbert bi-module is a left $\cal A$ and a right $\cal B$ module that fulfills
certain natural algebraic relations. Morita equivalence of $\cal A$ and $\cal
B$ is then equivalent to the existence of such a Hilbert bi-module.

To define a field theory on a space, one should first identify the appropriate
description of the fields. In the commutative case, this is that of sections
in bundles over the base space. The non-commutative analog of a bundle is a
projective module (that is a module of finite rank that can be completed by a
direct sum with an other module to a free module).

An important property of Hilbert bi-modules is that they can be used to turn
projective $\cal B$ modules into projective $\cal A$ modules: Let $P$ be a
projective $\cal B$ module and $H$ a Hilbert bi-module. Then, 
$$H \otimes_{\cal B} P$$
is a projective $\cal A$ module. Thus, every field on a space described by
$\cal B$ can be translated to a field in the $\cal A$ description of the same
space via the existence of a Hilbert bi-module.

We have seen that a bi-module that relates two algebraic descriptions of some
space can also be used to translate the algebraic analogs of vector bundles,
that is fields between the two descriptions. Under this mapping the ``form''
of the field will not be preserved generically. In the following section, we
will give an example of a scalar field that is mapped to a matrix valued field
and the change of the rank of the gauge group encountered in stringy
realizations of gauge theories on non-commutative spaces is yet another
manifestation of this phenomenon.

From a more abstract point of view, Morita equivalence of two algebraic
descriptions of one space maps the set of all field theories on that algebra
to the set of field theories on the other algebra rather than the pairs one
might naively expect: The scalar field on one algebra is not necessarily
mapped to the scalar field but to a scalar field tensored with the bi-module
which generically introduces matrix structure.

\chapter{Relating field theories on Morita equivalent descriptions of a space}
In the previous chapter we have mapped the fields, the constituents of a
physical theory.
It remains to translate the dynamics of the field theories between the two
settings. This can, for example, be done by translating action
functionals. Rather than continuing within the abstract setting, we will now
exemplify this in the concrete example of the non-commutative torus with
parameter $\theta$ for which the Morita equivalence classes are given by the
$SL(2,\MZ)$ orbits in the space of $\theta$'s. We will follow \cite{GT}, see
also \cite{AMNS}.

The $N\times N$ 't Hooft clock
and shift matrices
$$Q=\pmatrix{1&&&\cr
&e^{2\pi i\over N}&&\cr
&&\ddots&\cr
&&&e^{2\pi i(N-1)\over N}\cr},\qquad
P=\pmatrix{0&1&&\cr
&0&1&\cr
&&\ddots&\ddots\cr
1&&&\cr}.\gln\PQdef$$
obey the commutation relations 
$$PQ=QPe^{2\pi i\over N}\gln\commrel$$
and generate the algebra of complex $N\times N$ matrices as all $P^nQ^m$ form
a basis for $n,m=0,1,\ldots,N-1$.

We are interested in fields $\phi$ that take values in this algebra and live
on a two-torus with both radii equal to $R$. Let us demand that $\phi$ obeys
the twisted boundary conditions
$$\eqalign{\phi(x_1+2\pi R,x_2) &= P^{-m}\phi(x_1,x_2)P^m\cr
\phi(x_1,x_1+2\pi R) &= Q\phi(x_1,x_2)Q^{-1}\cr}\gln\bndcond$$
for some given integer $m$, later referred to as the magnetic flux. We assume
that $m$ and $N$ are relatively prime and fulfill
$$aN-cm=1$$ 
(or $c\equiv m^{-1} (N)$) for some integers $a$ and $c$. Note that
$\phi^\dagger$ fulfills the same boundary conditions.

We decompose $\phi$ into its Fourier-modes as
$$\phi(x_1,x_2)=\sum_{\vec r}\phi_{\vec r} Q^{-cr_1}P^{r_2} \exp\left(-i\pi\br
cNr_1r_2\right) \exp\left(-i {\vec r\cdot \vec x\over NR}\right)\gln\modes.$$
The first phase factor could be absorbed in the definition of $\phi_{\vec r}$
but we put it here for later convenience.

Using these modes, we can now define a complex scalar field $\hat\phi$ that
lives on a larger torus with radii $R'= NR$ as
$$\hat\phi(x_1,x_2) = \sum_{\vec r}\phi_{\vec r}\exp \left(-i{\vec r\cdot\vec
x\over R'}\right).$$

As can be checked by direct calculation, the operation ``$\hat{\phantom\phi}
$'' has the properties
$$\eqalign{(i)\qquad \widehat{\partial_i \phi} &= \partial_i\hat\phi\cr
(ii)\qquad \br 1{N^2}\int_{T^{2}{}'}\!\!\! d^2 x\, \hat\phi &=
\int_{T^2}\!\!\! d^2 x\, \tr \phi\cr
(iii)\qquad \widehat{\phi_1\phi_2} &= \hat\phi_1 * \hat\phi_2,\cr}$$
where in the last line we introduced the Moyal product
$$(\phi_1 * \phi_2) (\vec x) = e^{i{\vartheta^{ij}\over 2}{\partial\over\partial
x^i}{\partial\over\partial y^j}}\phi_1(\vec x)\phi_2(\vec y)|_{x=y}$$
with non-commutativity parameter $\vartheta^{ij}=\vartheta
\epsilon^{ij}=-2\pi\br cN {R'}^2 \epsilon^{ij}$. 

From these properties it follows that any action $\cS$ of a field theory in
terms of $\phi$'s on $T^2$ is equivalent to a dual action $\hat\cS$ in terms
of the $\hat\phi$'s if all products are replaced by Moyal products. In one
theory the non-commutativity lies in the matrix product in the other theory
non-commutativity comes with the Moyal product.

The two dimensions of the tori need not be the only dimensions of
space-time. All fields $\phi$ and $\hat\phi$ and therefore the modes
$\phi_{\vec r}$ could depend on any number of further uncompactified
directions. 

The classical observables of both theories can be translated to expressions of
the modes $\phi_{\vec r}$ and are therefore in a one-to-one
correspondence. Of course, the localization of the same observable in the two
theories is quite different, thus the question arises if it is possible to map
observables between the two theories without knowing about the relation
between the two mode expansions. 

The puzzle is that, in the hatted theory, $\vartheta$ should be thought of as a
real parameter, as a similar parameter arises as a background field in
low-energy effective theories arising from string theory. On the other hand,
the first model does not make sense for irrational values of 
$$\theta= {\vartheta\over 2\pi {R'}^2}$$
and appears very different for close real values as $1/2$ and
$1000000/2000001$. 

It would be nice to identify properties (observables?) in the theories for
similar values of $B$ that would show, that these theories are also similar in
some sense. A similar question was addressed in \cite{GT}. There, Wilson-loops
were considered for gauge theories that are related in same fashion as above
without a conclusive answer. Here, we would like to point out, that the same
question arises in a much larger class of theories and should therefore have
an answer beyond the consideration of Wilson loops. 

Even more, this question should be easier to analyze in a theory without
unphysical (gauge) degrees of freedom as this one where observables can be
directly identified.

What is behind this discussion is of course Morita equivalence: There is a
$SL(2,\MZ)$ action on $\theta$ as
$$\rho_{\pmatrix{a&b\cr c&d\cr}} (B) = {a\theta+b\over c\theta+d}$$
and the above relation is an explicitly
applied Hilbert-bi-module. As the algebras 
$${\cal A}= \{f\colon T^2\to \MC, \hbox{pointwise product}\},\qquad
{\cal B} = \{f\colon T^2{}'\to \MC, *\hbox{-product}\}$$
are Morita equivalent. Thus they have the same representation theory and
describe the same space from the perspective of non-commutative geometry. One
would expect to find a one-to-one correspondence between field-theories on
both spaces. This correspondence is spelled out in this note. 

What we have seen is that a theory that appears to be a theory on a
non-commutative space can be physically equivalent to theory of (possibly
large) matrices on a commutative space with magnetic background flux. As in
this realization the size of the matrices and the amount of flux depends
non-continously on $\theta$ it seems very unlikely that all physical theories
on these spaces (keep in mind this transformation can be applied to any
Lagrangian) share some ``magic'' property that nevertheless realizes physics
continously given the vastly different realizations on commutative spaces.

\chapter{Solving equations on the non-commutative torus}
We would like to investigate the dependence of physical quantities on
$\theta$, the parameter of non-commutativity. As explained before, we are
especially interested in whether physics is continuous in $\theta$.  Ideally,
one would like to calculate for example correlation functions of some quantum
field theory and study their behavior as $\theta$ varies. Unfortunately, this
is presently beyond our power. A more modest goal would be to solve classical
field equations on a non-commutative torus and determine whether solutions
depend continuously on $\theta$. Instead, here we even ask a simpler question:
Can we solve ``algebraic''\footnote{$^*$}{Of course, the $*$-product introduces
derivatives and the equation is really a partial differential equation.}
equations. 

The simplest example of this kind is: Find smooth functions $f\colon
T_\theta\to\MC$ that solve
$$f*f=f.\gln\projector$$
This equation for the non-commutative torus has been solved in terms of
$\vartheta$-functions in \cite{B}\ and was discussed in \cite{MM}, \cite{BKMT},
and \cite{KS}.
This equation is easily solved in the operator formalism that amounts to
replacing functions $f$ subject to *-multiplication by operators $\O f$
on a Hilbert space that implement the *-product via $\O {f*g}=\O f\O g$. The
inverse operation to this Moyal-Weil correspondence is 
given by 
$$f(x,y) = \Tr\left(\O f\int\!\!{dk\,dl\over (2\pi)^2} e^{ik\O x+il\O y}
e^{-ikx-iky}\right).$$ 
In the operator language, \projector\ just states that $f$ is a projector. Let
us assume it has rank one, that is $\O f=|\psi\rangle\langle\psi|$. Let us
translate this back into the language of *-products. For convenience, we use
an $x$-space basis $\{|x\rangle : x\in\MR\}$ for the Hilbert space, that is
$\O x|x\rangle = x|x\rangle$ and $\langle x|\psi\rangle=\psi(x)$:
$$\eqalign{f(x,y)&=
\Tr\left( |\psi\rangle\langle\psi|\int\!\!{dk\,dl\over
(2\pi)^2} e^{ik\O x+il\O y} e^{-ikx-iky}\right)\cr
&= \int\!\! dz\,\langle z|\psi\rangle\int\!\!{dk\,dl\over (2\pi)^2}
\langle\psi|e^{ik\O x}e^{il\O y}e^{-\br 12 [ik\O x,il\O y]}|z\rangle\,
e^{-ikx-iky}\cr 
&= \int\!\! dz\int\!\!{dk\,dl\over (2\pi)^2} \psi(z)\,\psi^*(z+l\theta)\,
e^{ik(z+\br 12\theta l-x)} e^{-ily}\cr
&=\int\!\! {dl\over 2\pi}\, \psi(x-\br 12\theta l)\,\psi^*(x+\br 12\theta
l)\,e^{-ily}\cr 
}$$
Here we used the Baker-Campbel-Hausdorff formula and the fact that the operator
$e^{il\O y}$ operates by shifting $x$ by $l\theta$.

So far, the calculation is equally valid on the non-commutative plane. Now, we
have to make use of the fact that $f(x,y)$ is supposed to be a function that
is defined on the torus. That is, it should be periodic in both $x$ and $y$
with unit period, say. This implies restrictions on the wave function
$\psi(x)$. Periodicity in $x$ forces $|\psi|$ to have unit periodicity, too. On
the other hand, $\psi(x-\br 12\theta l)\,\psi^*(x+\br 12\theta l)$ is the
Fourier transform in $y$. In order to have unit periodicity in $y$, this
Fourier transform must have support only for integer $l$.


We can make this even clearer if we decompose $\psi$ into its Fourier
modes. Naively, one would take
$$\psi(x) = \sum_{n\in\MZ} a_n e^{inx}.\gln\toonaive$$
This gives
$$\eqalign{f(x,y)&=\int\!{dl\over 2\pi}\sum_{m,n} a_m a_n^*
e^{i(m-n)x}\,e^{-il[(m+n)\theta +y]}\cr
&=\sum_{m,n} a_ma_n^* e^{i(m-n)x} \delta((m+n)\theta+y).\cr}$$
But we see that $x$ appears only thru $\exp(i(m-n)x)$. Thus, $m$ and $n$ do
not need to be integer moded but only their difference has to be in order to
ensure periodicity in $x$. Therefore we should replace \toonaive\ by 
$$\psi(x) = \sum_{n\in\MZ} a_n e^{i(n+\phi)x}$$
for some $\phi$ and obtain
$$f(x,y)=\sum_{m,n} a_ma_n^* e^{i(m-n)x}
\delta((m+n+2\phi)\theta+y).\gln\finalproj$$  
Now, this expression has to be periodic in $y$. That means, if it is non-zero
for some $y_0$ (meaning $(m+n+\phi)\theta=-y_0$ for some $m$ and $n$) it also
has to be for $y_0+1$. This can only be if for some other $m'$ and $n'$ we
have $(m+n-m'-n')\theta =1$. Finally, we find $\theta=1/q$ for integer $q$
especially, it cannot be irrational.

In order to find solutions for \projector\ if $\theta=p/q$ one must generalize
the above construction to the case of rank $p$ projectors. In that case, one
starts with $p$ orthogonal wave functions
$$\psi_i(x) = \sum_{n} a_{in} e^{i(n+\phi_i)x}$$ 
with ``shift angles'' $\phi_i=\phi_0+i/2q$ that fill up the ``missing'' peaks
for unit periodicity in $y$. This fits well with the intuition from string
theory: There, the  non-commutative torus arises from compactifications on a
torus in which a gauge field has a flux $\theta$ thru the torus. This setup
however is T-dual to a D1-brane that wraps $p$ and $q$ times the two basic
holonomy cycles of the torus. Thus, above each $x$ there are $p$ instances of
the brane. On the other hand, the rank of the projector has been argued to be
related to the number of branes in the context of non-commutative
instantons\cite{HKLM}. 

For irrational $\theta$, one could superimpose an infinite number of D-branes
densely. This yields a projector of infinite rank that is not well described
in the formalism we employ here. Those projectors have been constructed in
\cite{B}\ in terms of $\vartheta$-functions, see also \cite{MM}\ for a
discussion. 

So far in this section, we have concentrated on the distinction between
rational and irrational $\theta$. But of course we should also come back to
the question of the preceeding section of whether physical quantities are
similar for similar values of $\theta$ like $1/2$ and
$1000001/20000000$. As explained above, we intend to model a classical field
equation with \projector. In fact, such projectors have been used heavily to
construct soliton like solutions in non-commutative field theories in the
limit of vanishing derivatives (usually called limit of large
non-commutativity, but as we show here, it does not really make sense to talk
about limits of $\theta$ on compact spaces) as pioneered by \cite{GMS}. For
classical field theories without gauge invariances the observables the values
of the fields and their derivatives at points in space-time.
 
Once again, form the structure of \finalproj, it is clear
that possible functions $f(x,y)$ differ significantly since the denominator
of $\theta$ already determines the periodicity. From this we can conclude
that there is no notion of continuity in $\theta$ on the torus and
consequently limits with respect to $\theta$ cannot be defined as already
claimed several times.

One might wonder whether this behavior is a peculiarity of the classical
theory and it goes away once the theory is quantized. To us, thus sounds
highly unlikely although we do not have a proof for this (mainly due to the
lack of reliable information non-perturbative quantum observables of such
theories). At least we can comment on the ``$\theta$ enters only as a
parameter in the vertices'' argument. First of all, already this should also
apply to the classical theory if in the calculation, one includes tree-level
diagrams only. The conclusion of this note is to show that this cannot be
true. So, what is the fallacy?

By using this argument one forgets that it should be understood in a
path-integral context. In the non-commutative setting considered here, one
should be careful in choosing the correct set of functions to integrate
over. Namely, one should only integrate over functions that are compatible
with the periodicity of the non-commutative torus. In fact, this is what
stands behind our reasoning: The non-locality of the $*$-product makes the
issue of periodicity subtle and non-continuous in $\theta$ and thus the set of
allowed functions depends non-trivially on $\theta$. This resolves the
apparent puzzle that naively $\theta$ enters Feynman rules only as a parameter
in the vertices.

Let us end this section by giving an example of a system that behaves
discontinuously with respect to a parameter just as the non-commutative
torus. It consists of a field $\phi$ obeying the two dimensional wave-equation
$(\partial_t^2-\partial_x^2)\phi=0$. But instead of giving initial values for
$\phi$ and $\dot\phi$, we put the field in a rectangular, $a\times b$ sized,
space-time box and impose Dirichlet boundary conditions. For example we
require the field to vanish for $x=0$ or $x=a$ or $t=0$ or $t=b$. See
fig.~1.\epsfxsize=4cm \psfig{The
field $\phi$ in a space-time box}{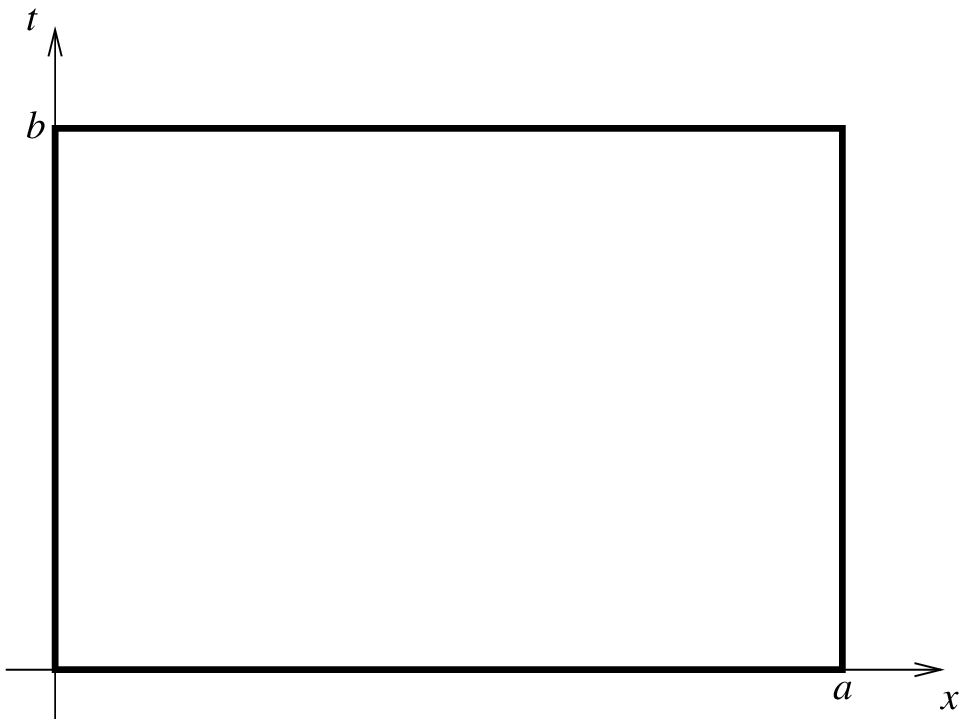}

We solve the field equation in terms of left- and right-movers as
$\phi(t,x)=u(x+t)+v(x-t)$ in the bulk and impose the boundary conditions. From
$\phi(0,x)=0$ we find $u=-v$, form $\phi(t,0)=0$ we find $u(t)=-u(-t)$. Hence,
the solution is determined in terms of one antisymmetric function. Furthermore 
$\phi(b,x)=0$ implies $u(x+b)=u(x-b)$, thus $u$ is $2b$-periodic. On the other
hand, from $\phi(t,a)=0$ follows $u(t+a)=u(t-a)$, so $u$ is also
$2a$-periodic. Whether this  double periodicity can be fulfilled for other
functions than $\phi(t,x)=0$ depends on the dimensionless ratio $\eta=a/b$: If
$\eta$ is irrational, the trivial one is the only smooth solution (as an aside
we note that if we allow for discontinuous solutions, the remaining freedom of
$u$ is precisely to pick a function on the irrational foliations of a
torus. But this space of foliations is described by a non-commutative torus
with irrational $\theta$, see \cite{C}).

If the ratio is rational $\eta=\br pq$ the periodicity depends on nominator
and denominator separately; it is given by $\br ap=\br bq$. The solution is a
linear combination of functions
$$u(x+t)=-v(x-t)=\sum_n b_n\sin \left(\pi npx\over a\right).$$ 
Just as for the projectors on the non-commutative torus we have a moduli space
of solutions that depends very critically on the dimensionless ratio
$\eta=a/b$. The connection between the two problems is that also in the case
of the non-commutative torus the *-product implies relations among the values
of the solution at points that a separated by $\theta$ and the periodicity
in $x$ implies relations for points at unit distance.

\chapter{The non-commutative torus and the non-commutative plane}
In what we have said above about the non-commutative torus it was crucial that
the torus is compact. In general, the background field $\vartheta^{ij}(x)$
that describes the non-commutativity of the space via
$$[x^i,x^j]=i\vartheta^{ij}(x)$$
has dimension of length squared. For a dimensionful quantity, of course, there
is no notion of rational or irrational as it would depend on the units used. 

Therefore, in infinite space we would not expect any effects as explored in
the previous two sections. Using string theory to realize field theories on
the non-commutative plane, $\vartheta$ is really just a background field on
which physics depends continuously. In perturbative calculations it only
enters as a parameter in the vertices and as such can be thought of as a
coupling constant.

In contrast, on compact spaces, there is the volume of the space that can be
used to form a dimensionless number $\theta$ from $\vartheta$ and it is this
dimensionless number that we have used in our discussion as we fixed our units
by requiring the torus to have unit periodicity. More precisely, from a
stringy perspective, $\theta$ is related to the flux of the $B$-field or a
gauge field thru the torus. For such a flux a quantization does not come
unexpected. In fact, multiplied by the correct electric charge, a rational
flux is integer in the correct normalization\cite{BGKL}.

This can also be viewed from a T-dual perspective: The space on which the
field theory lives is really a stack of D2-branes wrapping the torus. Now, we
can apply T-duality in one of the directions. This turns the D2-branes into
D1-branes wrapping the other torus direction. The non-commutativity $\theta$
can be attributed to a gauge field flux $F$ before the duality map is
applied. 

\epsfxsize=4cm\psfig{A D1 with
$\cot\alpha=\theta=\br 23$ and a finite mass string}{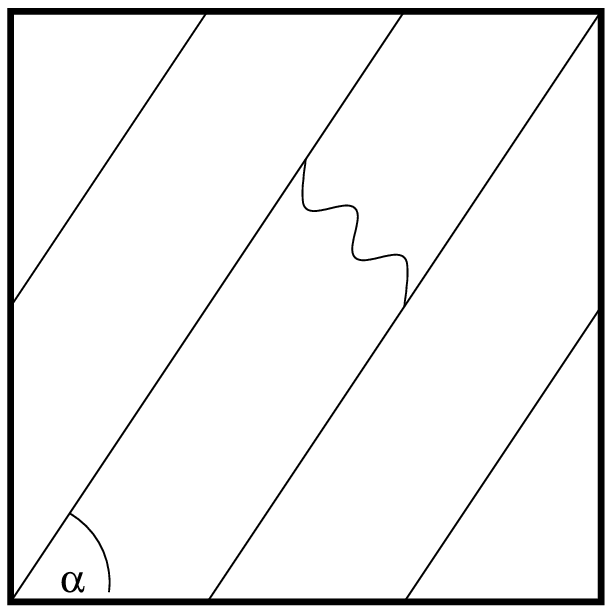}
Under T-duality, this flux is mapped to the slope of the D1, 
the angle being given by $\cot\alpha = F$. If $F$ (and thus $\theta$) is
irrational the D1 never his itself again and wraps the torus densely. There is
an infinity of arbitrary light open strings that span between different but
close wrappings that will dominate the dynamics. If $F$ is rational ($p/q$
say) the D1-brane will hit itself after wrapping the torus $q$ times and the
distance (and thus the mass of the strings stretching) of the different
wrappings stays finite. 

Thus again we find totally different behavior in the two cases which once
again leads us to the conclusion that physics on the torus cannot be
continuous with respect to $\theta$. 

Instead of the gauge field $F$ we could have also considered the NS two-form
$B$ since they only appear in the gauge invariant combination ${\cal
F}=B-F$. The geometric picture in that case differs as $B$ does not appear as
the slope of the D1-brane but rather as the tilt of the T-dual torus: Together
with the volume it forms the K\"ahler modulus $\rho=B+iV$ of the torus that
gets exchanged with the complex structure modulus $\tau$ upon
T-duality.\epsfxsize=6cm\psfig{The tilted torus with a ``non-local'', winding
string}{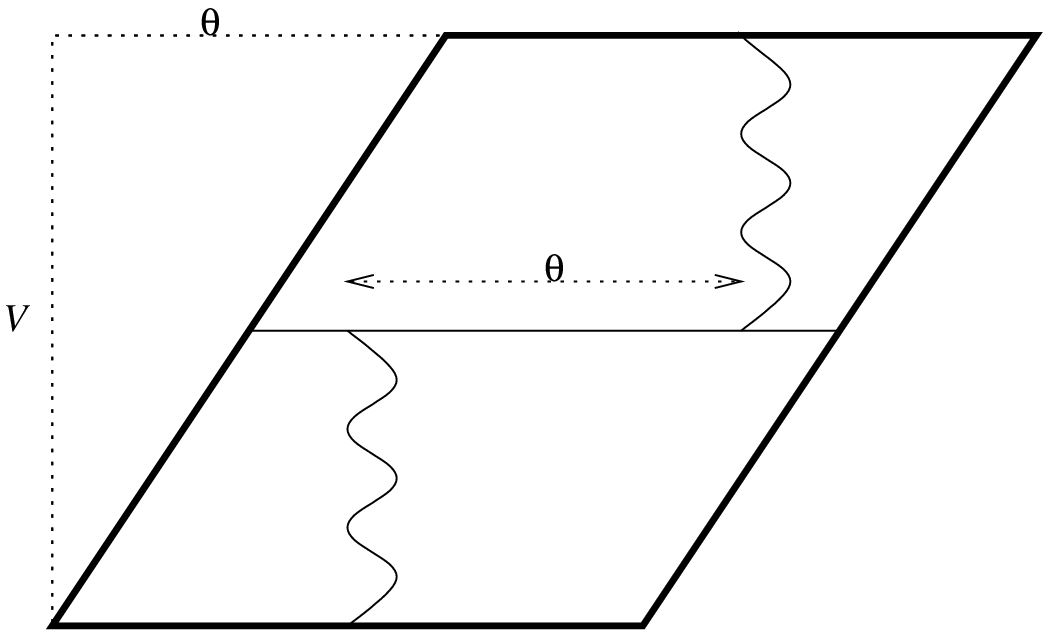}  

Indeed, one can see that the winding strings introduce a non-locality that
connects points on the D1-brane that are apart by integer multiples of
$\theta$ just as in \finalproj. The resulting theory in the $V\to 0$ limit on
$\MR/\MZ\times\theta\MZ$ is once again the theory on the non-commutative torus
as described by Douglas and Hull\cite{DH}.

\chapter{Conclusions}
In this note we have studied field theories on non-commutative tori. We paid
special attention to their behavior under variations of $\theta$ the
parameter of non-commutativity. 

We explained that Morita equivalence of spaces induces relations between field
theories living on these spaces that imply relations between theories for
different $\theta$ and varying magnetic flux.

Especially we constructed solutions to the projector equation and found
non-trivial dependence on $\theta$. We argued that it is likely that this is
the generic behavior of field equations on compact non-commutative spaces. 
Therefore, taking limits with respect to $\theta$ seems ill-defined. Likewise,
the IR-limit that uncompactifies the torus should be taken with care as the
compact and non-compact cases are qualitatively different. It would also be
interesting to study this in the light of the UV-IR-correspondence of
\cite{MvRS}\ that transforms UV singularities of a theory into IR singularities
on the non-commutative plane. As Morita equivalence can be used to transform
non-commutative tori to commutative tori with flux, this relation will be more
involved on compact spaces. We hope to come back to this question in the
future.

\closebib
\chapter{References}
\bigskip
\parindent =2cm
\eprint{AGB}{Alvarez-Gaume, L. and Barbon, J. L. F.}{Morita duality and large-N limits}{arXiv:hep-th/0109176}{}{Cited: 3 }
\paper{AMNS}{Ambj\o rn, Jan and Makeenko, Y. M. and Nishimura, J. and                  Szabo, R. J.}{Lattice gauge fields and discrete noncommutative Yang-Mills                  theory}{JHEP}{05}{2000}{023}{{\tt arXiv:hep-th/0004147} }{Cited: 5 }
\paper{B}{Boca, Florin B.}{Projections in Rotation Algebras and Theta Functions}{Cinnun. Math. Phys.}{202}{1999}{325---357}{{\tt } }{Cited: 8 9 }
\paper{BGKL}{Blumenhagen, Ralph and G\"orlich, Lars and K\"ors, Boris and                  L\"ust, Dieter}{Noncommutative compactifications of type I strings on tori                  with  magnetic background flux}{JHEP}{10}{2000}{006}{{\tt arXiv:hep-th/0007024} }{Cited: 11 }
\paper{BKMT}{Bars, I. and Kajiura, H. and Matsuo, Y. and Takayanagi, T.                  }{Tachyon condensation on noncommutative torus}{Phys. Rev.}{D63}{2001}{086001}{{\tt arXiv:hep-th/0010101} }{Cited: 8 }
\buch{C}{Connes, Alain}{Noncommutative Geometry}{1994}{Academic Press}{}{}{Cited: 2 11 }
\paper{CDS}{Connes, Alain and Douglas, Michael R. and Schwarz, Albert}{Noncommutative geometry and matrix theory: Compactification                  on tori}{JHEP}{02}{1998}{003}{{\tt arXiv:hep-th/9711162} }{Cited: 2 }
\paper{DH}{Douglas, Michael R. and Hull, Christopher M.}{D-branes and the noncommutative torus}{JHEP}{02}{1998}{008}{{\tt arXiv:hep-th/9711165} }{Cited: 2 12 }
\eprint{DN}{Douglas, Michael R. and Nekrasov, Nikita A.}{Noncommutative field theory}{arXiv:hep-th/0106048}{}{Cited: 2 }
\paper{GMS}{Gopakumar, Rajesh and Minwalla, Shiraz and Strominger,                  Andrew}{Noncommutative solitons}{JHEP}{05}{2000}{020}{{\tt arXiv:hep-th/0003160} }{Cited: 10 }
\eprint{GT}{Zachary Guralnik and Jan Troost}{Aspects of gauge theory on commutative and noncommutative                  tori}{hep-th/0103168}{}{Cited: 3 5 7 }
\paper{HKLM}{Harvey, Jeffrey A. and Kraus, Per and Larsen, Finn and                  Martinec, Emil J.}{D-branes and strings as non-commutative solitons}{JHEP}{07}{2000}{042}{{\tt arXiv:hep-th/0005031} }{Cited: 9 }
\paper{KS}{Krajewski, Thomas and Schnabl, Martin}{Exact solitons on noncommutative tori}{JHEP}{08}{2001}{002}{{\tt arXiv:hep-th/0104090} }{Cited: 8 }
\eprint{L}{Landi, Giovanni}{An introduction to noncommutative spaces and their                  geometry}{arXiv:hep-th/9701078}{}{Cited: 2 3 4 4 }
\eprint{MM}{Martinec, Emil J. and Moore, Gregory W.}{Noncommutative solitons on orbifolds}{arXiv:hep-th/0101199}{}{Cited: 8 9 }
\paper{MvRS}{Minwalla, Shiraz and Van Raamsdonk, Mark and Seiberg,                  Nathan}{Noncommutative perturbative dynamics}{JHEP}{02}{2000}{020}{{\tt hep-th/9912072} }{Cited: 13 }
\paper{S}{Schwarz, Albert}{Morita equivalence and duality}{Nucl. Phys.}{B534}{1998}{720--738}{{\tt arXiv:hep-th/9805034} }{Cited: 4 }
\paper{SW}{Seiberg, Nathan and Witten, Edward}{String theory and noncommutative geometry}{JHEP}{09}{1999}{032}{{\tt arXiv:hep-th/9908142} }{Cited: 2 }

\bye